\documentclass[aps,pra,prabib,twocolumn,showpacs,nofootinbib]{revtex4}
\usepackage{graphicx} \usepackage{amsmath} \usepackage{amssymb}
\usepackage{amsfonts} \usepackage{bm}

\begin{document}

\newcommand{\be}{\begin{equation}} \newcommand{\ee}{\end{equation}}
\newcommand{\bea}{\begin{eqnarray}}\newcommand{\eea}{\end{eqnarray}}

\title{Inequivalent quantization  in the field of a
ferromagnetic wire}

\author{Pulak Ranjan Giri} \email{pulakranjan.giri@saha.ac.in}

\affiliation{Theory Division, Saha Institute of Nuclear Physics,
1/AF Bidhannagar, Calcutta 700064, India}

\begin{abstract}
We argue that it is possible to bind neutral atom (NA) to the
ferromagnetic wire (FW) by inequivalent quantization of the
Hamiltonian. We follow the well known  von Neumann's method of
self-adjoint extensions (SAE) to get this inequivalent quantization,
which is characterized by a parameter
$\Sigma\in\mathbb{R}(\mbox{mod}2\pi)$. There exists a single bound
state for the coupling constant $\eta^2\in[0,1)$. Although this
bound state should not occur due to the existence of classical scale
symmetry in the problem.  But since quantization procedure breaks
this classical symmetry, bound state comes out as a scale in the
problem leading to scaling anomaly. We also discuss the strong
coupling region $\eta^2< 0$, which supports bound state making the
problem re-normalizable.

\end{abstract}

%\vspace{1cm}


\date{\today}

\maketitle

%%%%%%%%%%%%%%%%%%%%%%%%%%%%%%%%%%%%%
\section{Introduction}
%%%%%%%%%%%%%%%%%%%%%%%%%%%%%%%%%%%%

The quantum mechanics of a neutral atom (NA) or a neutral particle 
in the magnetic field
$\bf B$ created by a ferromagnetic wire (FW) \cite{tka} becomes a
nontrivial problem when the spin $\bf S$ of the neutral atom or the particle is
taken into account. The interaction of the spin with the magnetic
field generates a potential $V=-\boldsymbol{\mu}.\boldsymbol{B}$
($\boldsymbol{\mu}$ is the magnetic moment of the atom), which is
inverse square in nature for a certain alignment of the
magnetization of the wire. It is usually assumed that this system
does not have any stable bound state \cite{landau}
and depending on the sign of
the potential it is either unbounded in the field of the wire or it
falls to the center.

However, with the advance of research work on mathematical physics
we now know that systems with inverse square potential in
$1$-dimension \cite{reed}, $2$-dimensions \cite{giri1,giri},
$3$-dimensions \cite{camblong1,giri3} and even higher dimensions
\cite{kumar1} provide stable bound state solutions. It has been
possible to obtain bound state solutions due to the  consideration of
nontrivial boundary condition at the singularity of the Hamiltonian.
One possible way of constructing nontrivial boundary condition is to
start with a very restricted boundary condition and then go for a
possible SAE \cite{reed}. This method has been
very successfully implemented in different branches of physics
\cite{reed,giri1,giri,camblong1,giri3,kumar1,biru,
kumar2,kumar3,bh,stjep,giri4,feher}

In our present article we apply this well established method of
SAE to explain the quantum dynamics of a neutral
atom in the background magnetic field created by a ferromagnetic
wire. Although the mathematical technicalities needed for this
system is well know in the literature of mathematical physics, it is
still not very frequently used especially in molecular physics. In
our previous work we have used this same technique to explain the
formation of bound state solution of polarizable neutral atom in the
electric field of  charged single-walled carbon nano-tube (SWNT)
\cite{giri2}. It is argued that this bound state
could be a possible candidate for the smearing of the step edge of
quantized conduction  \cite{try}.

The article has been organized as follows: In Sec. \ref{FW}, we
review the system of neutral atom with spin  in the magnetic field
of a ferromagnetic wire. Due to the cylindrical symmetry, first the free
motion along $z$ direction has been separated out from the problem .
The remaining
two dimensional system has been reviewed and the radial eigenvalue
equation has been constructed, which will be analyzed in the next
section. In Sec. \ref{Hi}, Self-adjoint extensions (SAE) of the
radial eigenvalue equation has been made using the von Neumann's
method and bound state solution has been obtained. Re-normalization
techniques are discussed in Sec. \ref{Re}, to handle the strong
attractive, $\eta^2<0$, inverse square potential, which may arise in
our problem of neutral atom (NA) system. In Sec. \ref{Sc}, classical
scale symmetry of the full $3$-dimensional problem has been
discussed and the partial breaking of that classical scale symmetry
due to our quantization has been shown. The consequence of the
symmetry breaking is also discussed. Finally, we conclude in Sec.
\ref{Co}.

%%%%%%%%%%%%%%%%%%%%%%%%%%%%%%%%%%%%%%%%%%%%%%%%
\section{Neutral atom in the field of FW}\label{FW}
%%%%%%%%%%%%%%%%%%%%%%%%%%%%%%%%%%%%%%%%%%%%%%%%

We review here briefly about the  neutral atom system in the
magnetic field $\bf B$ of a ferromagnetic wire. The details can be
found in Ref. \cite{tka}. Let us consider a neutral atom of mass
$\mu$ moving in the magnetic field $\bf B$ of a ferromagnetic wire.
For the cylindrical symmetry of the system we consider the wire
along the $z$ axis.  The magnetization of the wire is considered
along the $x$ axis. The magnetic field is confined on the $x$-$y$
plane and  can be described in cylindrical coordinates $\rho, \phi,
z$ by $\boldsymbol{B}=\frac{2M}{\rho^2}(\cos 2\phi\hat i+\sin
2\phi\hat j)$ at a distance $\rho$ from the center of the wire. $M$
is the magnetization per unit length of the wire. The motion of the
atom along $z$ direction is a free particle motion, given by the
wave-function $e^{ikz}$ ($k$ is the wave-vector along the $z$
direction). We  therefore consider the $2$-dimensional problem on
$x$-$y$ plane. In  polar coordinates $(\rho,\phi)$, then the time
independent Schr\"{o}dinger equation for neutral atom system is of the form
($\hbar=2\mu=1$)
\begin{equation}
\left(-\nabla^2 - \boldsymbol{\mu}.\boldsymbol{B}-E\right)\Psi= 0\,,
\label{schrodinger}
\end{equation}
where $E$ is the energy eigenvalue of the neutral atom, the magnetic
moment $\boldsymbol{\mu}= -g\mu_0\boldsymbol{S}/\hbar$, $\bf
S$ is the spin of the atom, $g$ is the Lande factor and $\mu_0$ is
the Bohr magneton. Eq. (\ref{schrodinger}) is completely
separable and after substituting  $\Psi(\rho,\phi,z) = (1/\sqrt
\rho)R(\rho)\exp(il_z\phi)\chi$ into the Schr\"{o}dinger equation
(\ref{schrodinger}), the radial equation can be written in the
following well known $1$-dimensional eigenvalue problem with inverse
square interaction:
\begin{eqnarray}
\left(H_\rho-E\right) R(\rho) = 0\,, \label{radial}
\end{eqnarray}
where the radial Hamiltonian is given by $H_\rho=
-\frac{d^2}{d\rho^2}+\left(\eta ^2-1/4\right)/\rho^2$. The coupling
constant $\eta^2$ of this Hamiltonian $H_\rho$ can be obtained  by
solving the spin part \cite{tka},
\begin{eqnarray}
\left[(l_z-2S_z)^2-\gamma S_x-\eta^2\right]\chi=0\,.
\label{angularequnation}
\end{eqnarray}
Here we are not interested in spin part, because we concentrate
mostly on inequivalent quantization of the radial Hamiltonian. We
just state the results, which we  need in our discussion. $\eta^2=
l_z^2+1\pm \sqrt{4l_z^2+(\gamma/2)^2}$ for $S=1/2$ \cite{tka}, where
$\gamma= 2g\mu_0M$.  In order to
solve (\ref{radial}) we need to define a domain so that the
Hamiltonian $H_\rho$ becomes self-adjoint. In the next section we
perform the SAE of the Hamiltonian $H_\rho$ and from that we
determine the bound state eigenvalue.
\begin{figure}
%\begin{center}
\includegraphics[width=0.4\textwidth, height=0.2\textheight]{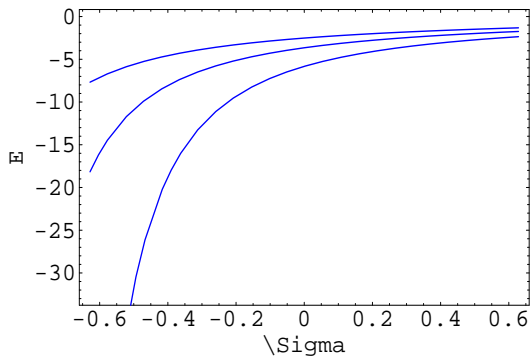}
\caption {(color online) A plot of the bound state energy $E$(dimension
  $length^{-2}$) as a function of
  the self-adjoint extension parameter $\Sigma$.
From top to bottom $\eta=0.7, 0.6, 0.5$ respectively.}
%\end{center}
\end{figure}

\begin{figure}
%\begin{center}
\includegraphics[width=0.4\textwidth, height=0.2\textheight]{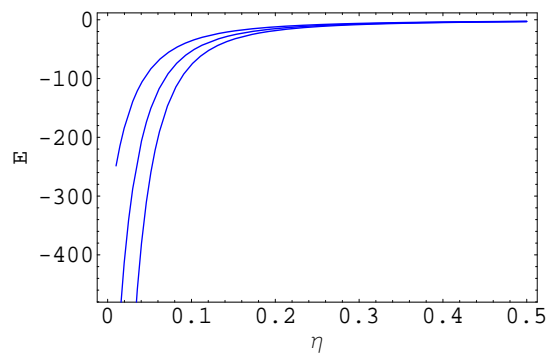}
\caption {(color online) A plot of the bound state energy $E$(dimension $length^{-2}$) as a function of
  the coupling constant $\eta\neq 0$.
From top to bottom $\Sigma=\pi/6,\pi/7,\pi/8$ respectively.}
%\end{center}
\end{figure}

%%%%%%%%%%%%%%%%%%%%%%%%%%%%%%%%%%%%%%%%%%%%%%%%%%
\section{SAE of the radial Hamiltonian}\label{Hi}
%%%%%%%%%%%%%%%%%%%%%%%%%%%%%%%%%%%%%%%%%%%%%%%%%

We need to construct a domain for our Hamiltonian $H_\rho$, because
otherwise the Hamiltonian does not have any meaning. $H_\rho$ is
formally self-adjoint, but formal self-adjointness of an operator
does not mean that it is self-adjoint. To start with, we search for a
very restricted domain $\mathcal{D}(H_\rho)$ so that $H_\rho$ is
symmetric (or hermitian) in that domain. $H_\rho$ can be made
symmetric for $\forall\psi_1,\psi_2\in\mathcal{D}(H_\rho)$, if the
R.H.S of
\begin{eqnarray}
\nonumber(H_\rho\psi_1(\rho),\psi_2(\rho))-
(\psi_1(\rho),H_\rho\psi_2(\rho))=\Delta(\infty)-\Delta(0),\label{symmetric}
\end{eqnarray}
where $\Delta(r)= \psi_1(\rho)\dot{\psi}_2(\rho)
-\dot{\psi}_1(\rho)\psi_2(\rho)$, is zero. The asymptotic limit of
the functions $\psi_1,\psi_2$ are assumed to fall to zero,  so we get
 $\Delta(\infty)=0$. The condition for symmetric Hamiltonian thus
reduces to $\Delta(0)=0$ and it can be easily achieved if the
elements of $\mathcal{D}(H_\rho)$ and their derivatives with
respective to $\rho$ becomes zero at origin. We then need to get the
domain $\mathcal{D}(H^*_\rho)$ of the operator $H^*_\rho$ (this
operator is adjoint to the Hamiltonian $H_\rho$). Since $H_\rho$ is
formally self-adjoint, its adjoint $H^*_\rho$ should have the same
form and the elements $\xi\in \mathcal{D}(H^*_\rho)$ can now be
found from
\begin{eqnarray}
(H^*_\rho\xi(\rho),\psi(\rho))-
(\xi(\rho),H_\rho\psi(\rho))=0,\forall\psi(\rho)\in
\mathcal{D}(H_\rho).\label{adjointequation}
\end{eqnarray}
We see that the elements $\psi(\rho)$
($\psi(\rho)\in\mathcal{D}(H_\rho)$) are so restricted that no
restriction on the elements $\xi(\rho)$ ($\xi(\rho)\in
\mathcal{D}(H^*_\rho)$) are required in order to satisfy
(\ref{adjointequation}). Since the two domains $\mathcal{D}(H_\rho)$
and $\mathcal{D}(H^*_\rho)$ are not equal, i.e.,
$\mathcal{D}(H_\rho)\neq \mathcal{D}(H^*_\rho)$, the operator
$H_\rho$ is not self-adjoint in the domain $\mathcal{D}(H_\rho)$. To
find out the possible SAE we follow the well
known von Neumann's method. We have to find out the square
integrable solutions of
%\begin{eqnarray}
$H_\rho^*\phi_{\pm} = \pm i \phi_{\pm}$.
%\label{def}
%\end{eqnarray}
The square integrable solutions can be written in terms of Hankel
functions ($H^{(1)}, H^{(2)}$) \cite{abr} as $\phi_+ (\rho) =
\sqrt{\rho}H^{(1)}_\eta (\rho e^{i \frac{ \pi}{4}})$ and $\phi_-
(\rho) = \sqrt{\rho}H^{(2)}_\eta (\rho e^{-i \frac{ \pi}{4}})$.
These solutions are square integrable at origin for
$\eta^2\in[0,1)$. This can be checked from the short distance
behavior
\begin{eqnarray}
\lim_{\rho\to 0}|\phi_\pm(\rho)|^2= ( )\rho+( )\rho^{2\eta+1} + (
)\rho^{-2\eta+1}\,, \label{short}
\end{eqnarray}
where $()$s are constants, which are unimportant for this purpose.
It can also be checked from (\ref{short}) that for $\eta^2\geq 1$
$\phi_\pm(\rho)$ are not square integrable at the origin. In this case
$H_\rho$ is essentially self-adjoint. Our next task is to get a
SAE for $\eta^2\in[0,1)$ and it will be characterized by a single
 parameter $\Sigma$. The
Hamiltonian $H_\rho$ will now be self adjoint over the newly defined
domain $ \mathcal{D}_\Sigma(H_\rho)\equiv D(H_\rho) + \phi_+(\rho) +
\exp(i\Sigma)\phi_-(\rho)$,
 %\label{ndomain}
 %\end{eqnarray}
 where $ \Sigma \in R$ (mod $2 \pi$) \cite{reed}.  Using
$\mathcal{D}_\Sigma(H_\rho)$ we have to calculate the bound state
solutions. The bound state eigenfunction of  (\ref{radial})  is
\begin{eqnarray}
\nonumber R(\rho)&\equiv& \sqrt{\rho}H^{(1)}_\eta(\sqrt{E}\rho),
~~~~~~~~~\eta \neq 0\\
&\equiv& \sqrt{-2E\rho} K_0\left( \sqrt{- E} \rho\right), \eta =0
\label{radial1}
\end{eqnarray}
where $K_0$ is the modified Bessel function \cite{abr}.
Note that for $\eta=0$ we just mention the results, but the calculations
should be done separately to get the results.
The bound state eigenvalue $E$ of (\ref{radial}) can be calculated from the
relation
\begin{equation}
\lim_{\rho\to 0}R(\rho)= \lim_{\rho\to 0}\left(\phi_+(\rho) +
\exp(i\Sigma)\phi_-(\rho)\right) \,. \label{relation1}
\end{equation}
Equating  the coefficients of $\rho^{\eta+1/2 }$ and $\rho^{-
\eta+1/2}$ from both sides of (\ref{relation1}) and comparing
between them we get a single bound state energy
\begin{eqnarray}
\nonumber E &=& - \sqrt[\eta]{ \cos(\pi\eta/2)+
\cot(\Sigma/2+\pi\eta/4)\sin(\pi\eta/2)}
    , \eta\neq 0\\
 &=& - {\exp}\left[(\pi/2) {\cot}(\Sigma/2)\right],
 ~~~~~~~~~~~~~~~~~~~~~~~~\eta=0
 \label{eigen1}
\end{eqnarray}
for a fixed value of $\eta^2$. Note that only those values of
$\Sigma$ give bound state solution for which the quantity under
third bracket in (\ref{eigen1}) is positive.  Since for different
values of the SAE parameter $\Sigma$ we get
different boundary conditions and thus different systems, this can
be recognized as inequivalent quantization.
\begin{figure}
%\begin{center}
\includegraphics[width=0.4\textwidth, height=0.2\textheight]{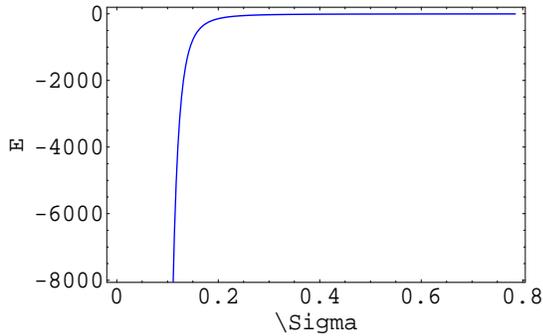}
\caption {(color online) A plot of the bound state energy
$E$(dimension $length^{-2}$) as a function of
  the self-adjoint extension parameter $\Sigma$ for $\eta=0$.}
%\end{center}
\end{figure}
In FIG. 1, the bound state energy $E$ has been plotted as a function
of the self-adjoint extension parameter $\Sigma$ for three different
values of the coupling constant $\eta\neq 0$. In FIG. 2, the same
bound state energy has been plotted as a function of the coupling
constant $\eta$ for three different values of the self-adjoint
extension parameter $\Sigma$. Since, for $\eta=0$, the expression
for the eigenvalue $E$ is different, it has thus been plotted in
FIG. 3 separately. The 3-dimensional plot of the eigenvalue $E$ as a function
of the two parameters $\eta$ and $\sigma$ is shown in FIG. 6.

%%%%%%%%%%%%%%%%%%%%%%%%%%%%%%%%%%%%%%%%%%%%%%%%%%%%%%%
\section{Re-normalization in NA system}\label{Re}
%%%%%%%%%%%%%%%%%%%%%%%%%%%%%%%%%%%%%%%%%%%%%%%%%%%%%%%

For strong attractive ($\eta^2<0$) inverse square potential, the
usual analysis of the NA system will give a tower of spectrum
\cite{case} with ground state being $-\infty$. Even the self-adjointness
technique \cite{feher} also gives the ground state to be -ve infinity. 
This implies that the
system will collapse if $\eta^2<0$. But re-normalization technique
\cite{rajeev,hor,bawin,braaten} has a remedy for this problem to
give a finite ground state, thus making the problem physically
realizable. In re-normalization technique, the divergent Hamiltonian
is regularized with an  ultraviolet cut off $\rho= \Theta$, for
example consider an  infinite barrier regularized potential
\begin{eqnarray}
\nonumber V(\rho)&=& \infty\,,~~~~~~~\mbox{for}~~\rho<\Theta \\
&=&\frac{\mbox{g}(\Theta)}{\rho^2}\,,
~~~\mbox{for}~~\rho\geq\Theta\label{renormalization1}
\end{eqnarray}
where now the coupling constant depends on the ultraviolet  cut off,
$\mbox{g}(\Theta)=(\eta^2(\Theta)-1/4)$. The ultraviolet cut off
allows the system to sustain a well defined bound state. This can be
understood form the regularized time independent Schr\"{o}dinger
equation
\begin{eqnarray}
\left(H_\rho(\Theta)- E(\Theta)\right)R(\rho,\Theta)=0\,,
~ \mbox{for}~\rho\geq\Theta
\label{regularized}
\end{eqnarray}
with the boundary condition that $R(\rho=\Theta,\Theta)=0$.
%Taylor series expansion of the L.H.S is of the following form
%\begin{eqnarray}
%\left(\nonumber \left(H_\rho(\Theta)- E(\Theta)\right)R(\rho,\Theta)\Big{|}_{\Theta=0}=0\right)+\\
%\Theta\left(\frac{d(\mbox{L.H.S})}{d\Theta}\right)\Big{|}_{\Theta=0}+
%\mathcal{O}(\Theta^2)=0 \label{regularized1}
%\end{eqnarray}
%In order to evaluate the beta function we need the the relation
The dependence of the coupling constant $\eta(\Theta)$  on the
ultraviolet cutoff $\Theta$ is encoded in the relation
\begin{eqnarray}
\frac{d H_\rho(\Theta)}{d\Theta}=
\left[\frac{d}{d\Theta},H_\rho(\Theta)\right]\,.
\label{regularized2}
\end{eqnarray}
The relation of the coupling constant with the ultraviolet cut off
can be obtained from (\ref{regularized2}) to be
\begin{eqnarray}
\frac{d\eta^2(\Theta)}{d\Theta}=
\rho^2\left[\frac{d}{d\Theta},H_\rho(\Theta)\right]\,.
\label{regularized3}
\end{eqnarray}
The coupling $\eta(\Theta)$ goes to zero in the limit $\Theta\to 0$
\cite{rajeev}. Thus $\eta=0$ is the ultraviolet stable point for the system.
\begin{figure}
%\begin{center}
\includegraphics[width=0.4\textwidth, height=0.2\textheight]{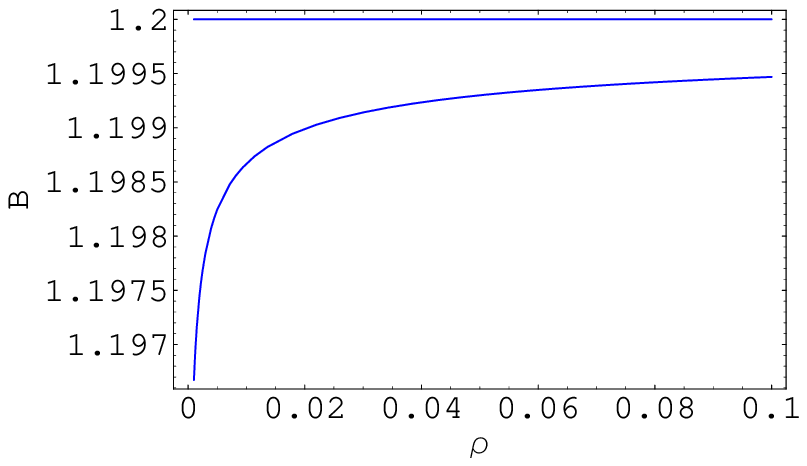}
\caption {(color online) A plot of the absolute value of the
quantity $B= \frac{|\lim_{\rho\to
0}\Lambda\phi(\rho)|}{|\lim_{\rho\to 0}\phi(\rho)|}$ as a function
of $\rho$ for $\eta=0.2$. The straight line corresponds to $\Sigma=
-\frac{\pi}{10}$ and curve line corresponds to $\Sigma=
-1.001.\frac{\pi}{10}$. The point $(\eta,\Sigma)\equiv
(0.2,-\frac{\pi}{4})$ associated with the straight line is the point
in the $\eta,\Sigma$ parametric space where scaling is unbroken on
the other hand at the point $(\eta,\Sigma)\equiv
(0.2,-1.001.\frac{\pi}{4})$ associated with the curve line, scale
symmetry is broken.}
%\end{center}
\end{figure}

\begin{figure}
%\begin{center}
\includegraphics[width=0.4\textwidth, height=0.2\textheight]{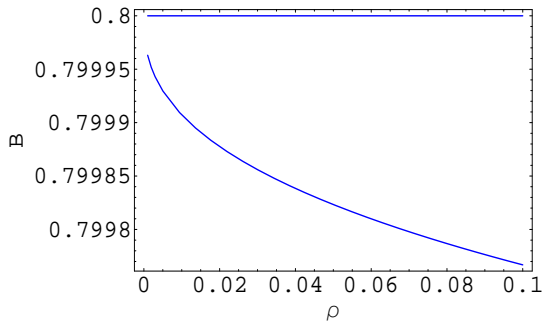}
\caption {(color online) A plot of the absolute value of the
quantity $B= \frac{|\lim_{\rho\to
0}\Lambda\phi(\rho)|}{|\lim_{\rho\to 0}\phi(\rho)|}$ as a function
of $\rho$ for $\eta=0.2$. The straight line corresponds to $\Sigma=
-\frac{3\pi}{10}$ and curve line corresponds to $\Sigma=
-1.001.\frac{3\pi}{10}$. The point $(\eta,\Sigma)\equiv
(0.2,-\frac{3\pi}{4})$ associated with the straight line is the
point in the $\eta,\Sigma$ parametric space where scaling is
unbroken on the other hand at the point $(\eta,\Sigma)\equiv
(0.2,-1.001.\frac{3\pi}{4})$ associated with the curve line, scale
symmetry is broken.}
%\end{center}
\end{figure}

%%%%%%%%%%%%%%%%%%%%%%%%%%%%%%%%%%%%%%%%%%%%%%%%
\section{Scaling anomaly of NA system}\label{Sc}
%%%%%%%%%%%%%%%%%%%%%%%%%%%%%%%%%%%%%%%%%%%%%%%%

We now discuss scaling symmetry and its anomaly
\cite{alfaro,esteve,gozzi,gozzi1,govinda,gino} for the full
$3$-dimensional problem of a neutral atom in the magnetic field of a
ferromagnetic wire. Classically the action constructed from  the
Hamiltonian $H={\boldsymbol p}^2/2\mu -
\boldsymbol{\mu}.\boldsymbol{B}$ is scale invariant under the scale
transformation $\boldsymbol{r}\to \varrho \boldsymbol{r}$ and $t\to
\varrho^2 t$, where $\boldsymbol{r}= x\hat i+y\hat j+z\hat k$,
$\varrho$ is the scale factor, $t$ is the time. The Hamiltonian
transforms as $H\to (1/\varrho^2)H$ under the scale transformation.
Lagrangian constructed from this Hamiltonian $H$ also transforms in
the same way $L\to (1/\varrho^2)L$. So the action of our system is
evidently scale invariant as stated above. The consequence of scale
invariance is that the system should not have any bound state.
Because, the presence of bound state energy would provide a scale
\cite{govinda} for the system and consequently  scale invariance
will break down.

We have seen in our previous section that after inequivalent
quantization of the Hamiltonian, the use of nontrivial boundary
condition gives a single bound state in the interval
$\eta^2\in[0,1)$ for the $2$-dimensional problem ($x$-$y$ plane).
This single bound state is however a characteristic feature of the
of the inverse square potential and has been obtained in literature
\cite{giri1,giri,giri3,giri4,alfaro,esteve,gozzi,gozzi1,govinda,gino,cabo}
before. Note that the motion along $z$ direction is still given by a
free particle solution $e^{ikz}$. Bound state energy given by
(\ref{eigen1}) provides a scale for the $2$-dimensional the system,
leading to ``partial scaling anomaly". We  use the term ``partial
scaling anomaly", because in the $z$ direction scaling symmetry is
still restored even after quantization. This situation happens in
other cylindrically symmetric systems also, for example the motion
of a charged particle or a dipole in cosmic string background
\cite{giri1,giri} show partial scale symmetry breaking. Quantum
mechanically scaling transformation is associated with a scaling
operator \cite{alfaro,govinda} $\Lambda = -\frac{i}{2} (\rho
\frac{d}{d\rho} + \frac{d}{d\rho} \rho)$. For a generic element
$\phi(\rho)\in \mathcal{D}_\Sigma(H_\rho)$ and for $\Sigma\neq
-\frac{ \eta \pi}{2}$ or $\neq -\frac{ 3 \eta \pi}{2}$, $\Lambda
\phi(\rho)\notin\mathcal{D}_\Sigma(H_\rho)$. This can be checked
from the relation
\begin{eqnarray}
\lim_{\rho\to 0}\Lambda \phi(\rho) \neq C\lim_{\rho\to 0}
\phi(\rho),~\mbox{for}~\eta\neq 0\label{scalingoperator}
\end{eqnarray}
where $C$ is any complex number. Since the action of $\Lambda$ on
the domain $\mathcal{D}_\Sigma(H_\rho)$ does not keep the domain
invariant, scaling symmetry is broken \cite{govinda}. But it can be
shown that for $\Sigma= -\frac{ \eta \pi}{2}$ and $-\frac{ 3 \eta
\pi}{2}$, scaling symmetry is restored  \cite{kumar1,kumar2,kumar3}
even after quantization. In this case
\begin{eqnarray}
\Lambda
\phi(\rho)\in\mathcal{D}_\Sigma(H_\rho),~~\mbox{for}~~\eta\neq 0
\label{}
\end{eqnarray}
For more clarity on scale symmetry and its anomalous breaking  we
plot the ratio of $|\lim_{\rho\to 0}\Lambda\phi(\rho)|$ to
$|\lim_{\rho\to 0}\phi(\rho)|$ as a function of the radial variable
$\rho$ for $\eta=0.2$. Note that for $\eta=0.2$, scale symmetry is
restored for $\Sigma= -\frac{\pi}{10}$ and $\Sigma=
-\frac{3\pi}{10}$ respectively. The straight line in FIG. 4
corresponds to $\Sigma= -\frac{\pi}{10}$ and straight line in FIG. 5
corresponds to $\Sigma= -\frac{3\pi}{10}$. On the other hand the
scaling anomaly is displayed by the two curved line in FIG. 4 and
FIG. 5 respectively. One can similarly discuss the scaling anomaly
for the $\eta=0$ case from its bound state solution and its domain.
It can also be noted that the two scaling symmetry points is
associated with two extreme points on the energy scale for bound
state of the system, one is at $E=0$ and other is at $E=\infty$.
\begin{figure}
%\begin{center}
\includegraphics[width=0.4\textwidth, height=0.2\textheight]{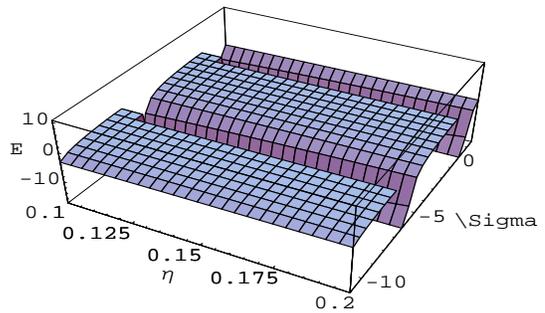}
\caption {(color online) A 3-dimensional plot of the eigenvalue $E$ as a
  function of the parameter $\eta$ and $\Sigma$.}
%\end{center}
\end{figure}

%%%%%%%%%%%%%%%%%%%%%%%%%%%%%%%%%%%%%%%%%%%%
\section{Conclusion}\label{Co}
%%%%%%%%%%%%%%%%%%%%%%%%%%%%%%%%%%%%%%%%%%%%

In literature \cite{tka} it is usually said that the neutral atom
would not have any stable bound state under the magnetic field of a
ferromagnetic wire and depending on the coupling constant $\eta^2$,
it will either be unbounded (for $\eta^2>0$) or fall into the
singularity (for $\eta^2<0$). In our work we have shown that the
assumption in the literature is not true, if we allow nontrivial
boundary condition to hold  for the system, contrary to the result
of the previous literature \cite{tka}, where usual boundary
condition has been considered. It is in fact possible to form stable
bound state for the $2$-dimensional problem in the field of a
ferromagnetic wire. The dynamics along $z$ direction is given by
free state solution  due to cylindrical  symmetry of the system. We
have also shown that scaling symmetry is partially violated due to
nontrivial quantization of the system. Scaling symmetry in $z$
direction still survives even after quantization, so there is no
bound state solution in $z$ coordinate. On the other hand on $x$-$y$
plane, scaling symmetry is broken due to quantization, leading to
bound state solution. For strong coupling region, $\eta^2<0$,
re-normalization technique can be used to find a physical bound
state solution. $\eta=0$ is then identified as the ultraviolet
stable fixed point for the neutral atom system.

%\vskip 0.5 cm

%\noindent {\bf Acknowledgment}


\begin{thebibliography}{99}

\bibitem{tka} V. M. Tkachuk, Phys. Rev. {\bf A60}, 4715 (1999).

\bibitem{landau} L. D. Landau and E. M. Lifshitz, {\it Quantum Mechanics},
(Pergamon Press, London, 1959).

\bibitem{reed} M. Reed and B. Simon,  {\it Fourier Analysis, Self-Adjointness}
  ( New York :Academic, 1975 ).

\bibitem{giri1} P. R. Giri, arXiv:0704.1725v2 [hep-th].

\bibitem{giri} P. R. Giri, Phys. Rev. {\bf A76},  012114 (2007).


\bibitem{camblong1} H. E. Camblong, L. N. Epele, H. Fanchiotti and
C. A. G. Canal, Phys. Rev. Lett. {\bf 87} 220402 (2001);

H. E. Camblong, C. R. Ordonez, Phys. Rev. {\bf D68}, 125013 (2003).

\bibitem{giri3} P. R. Giri, K. S. Gupta, S. Meljanac and A. Samsarov,
hep-th/0703121.

\bibitem{kumar1} B. Basu-Mallick, Pijush K. Ghosh and Kumar S. Gupta,
Nucl. Phys. {\bf B659}, 437 (2003).

\bibitem{biru} B. Basu-Mallick and Kumar S. Gupta, Phys. Lett. {\bf A292}, 36
(2001).


\bibitem{kumar2} B. Basu-Mallick, Pijush K. Ghosh and Kumar S. Gupta,
Phys. Lett. {\bf A311},  87 (2003).

\bibitem{kumar3} Kumar S. Gupta,  Mod. Phys. Lett. {\bf A18}, 2355 (2003).

\bibitem{bh} D. Birmingham, Kumar S. Gupta and Siddhartha Sen,
Phys. Lett. {\bf B505}, 191 (2001);

Kumar S. Gupta and Siddhartha Sen, Phys. Lett. {\bf B526}, 121 (2002).

\bibitem{stjep} S. Meljanac, A. Samsarov, B. Basu-Mallick and Kumar S. Gupta,
Eur.  Phys. J. {\bf C49}, 875 (2007).

\bibitem{giri4} P. R. Giri, arXiv:0708.0707v1 [hep-th].

\bibitem{feher} L. Feher, I. Tsutsui, T. Fulop,
Nucl.Phys. {\bf B715},  713 (2005).

\bibitem{giri2} P. R. Giri, arXiv:0706.2945v1 [cond-mat.mtrl-sci].


\bibitem{try} T. Ristroph, A. Goodsell, J. A. Golovchenko and L. V.  Hau,
Phys. Rev. Lett. {\bf 94}, 066102 (2005).

\bibitem{abr} M. Abromowitz, I. A. Stegun, {\it Handbook of Mathematical
Functions} (Dover, New York, 1970).


\bibitem{case} K. M. Case, Phys. Rev. {\bf 80}, 797 (1950).

\bibitem{rajeev} K. S. Gupta and S. G. Rajeev, Phys. Rev. {\bf D48}, 5940
(1993).

\bibitem{hor} H. E. Camblong, L. N. Epele, H. Fanchiotti and C. A. Garcia
Canal, Phys. Rev. Lett. {\bf 85}, 1590 (2000).

\bibitem{bawin} M. Bawin and S. A. Coon, Phys. Rev. {\bf A76}, 042712 (2003).

\bibitem{braaten} E. Braaten and D. Phillips, Phys. Rev. {\bf A70}, (2004).

\bibitem{alfaro} V. de Alfaro, S. Fubini and G. furlan, Nuovo
Cimento {\bf 34A}, 569 (1976).


\bibitem{esteve} J. G. Esteve, Phys. Rev. {\bf D66},   125013 (2002).


\bibitem{gozzi} E. Gozzi and D. Mauro, Phys. Lett. {\bf A345}, 273
(2005).

\bibitem{gozzi1} E. Gozzi and D. Mauro, J. Phys. {\bf A39}, 3411 (2006).

\bibitem{govinda} T. R. Govindarajan, V. suneeta and S. Vaidya, Nucl. Phys.
{\bf B583}, 291 (2000).

\bibitem{gino}  G. N. J. Ananos, H. E. Camblong, C. R. Ordonez,
Phys. Rev. {\bf D68},  025006  (2003).

\bibitem{cabo} A. Cabo, J. L. Lucio and H. Mercado,
Am. J. Phys. {\bf 66}, 240 (1998).

\end{thebibliography}
\end{document}